\documentstyle[12pt]{article}

\title{
Phonon-Induced Renormalization and Interaction: \\
An Improvement on Frohlich Transformation }

\author{H. Zheng\\
Department of Applied Physics, Shanghai Jiao Tong University,\\
Shanghai, P. R. China}

\date{}

\begin{document}

\maketitle

\begin{abstract}

{\baselineskip 20pt

\normalsize

Starting from the Flohlich electron-phonon model, which has a history
of 50 years, a new unitary transformation is proposed to implement 
the perturbation treatment. Our main results are:
(1)The phonon-induced interaction shows a crossover from the
BCS-like potential when the phonon frequency $\omega_p$ is much smaller
than the Fermi energy $E_F$ to that of the small polarons when
$\omega_p/E_F\ge 1$.
(2)The jump of momentum distribution of electron number
$n_{\bf k}$ at the Fermi surface goes to zero when the dimensionless
coupling constant $\lambda$ increases to the critical value
$\lambda_c\ge 1$, which means a possible broken down of the Fermi-liquid
description.

\vskip 1cm

}
\end{abstract}

\pagebreak

\baselineskip 20pt

It is a well-known fact that within the Migdal-Eliashberg (ME)
description\cite{mig,elia,sca,grim} of
electrons and phonons coupled by the linear electron-phonon interaction
there is no instability (where the Fermi-liquid description may break down)
at any value of the dimensionless coupling constant $\lambda$.
People tried to study the polaronic collapse of the electron band
starting from the Lang-Firsov (LF) transformation\cite{lf} followed by
the small polaron approximation\cite{hol,am}. Note that the LF transformation
together with the small polaron approximation cannot lead to those results
which can be obtained via the ME approach. For example, Bardeen, Cooper,
and Schrieffer\cite{bcs} (BCS) proposed a square-well potential
for electrons in momentum space, that is, only those electrons within
a layer of width $\omega_p$ (the characteristic phonon
frequency) near the Fermi surface can attract with each other. This
potential is localized in momentum space, so it is extended in real
space and leads to large coherence length of Cooper pairs. But the
LF transformation results in an attractive potential for all electrons
in Fermi sea\cite{lf,am}, which is a localized potential in real space.
As far as we know, there is no theories which can describe successfully
the crossover between the two pictures.
In this letter we propose a new approach which can (1)describe the
crossover and (2)lead to a possible broken down of the Fermi-liquid
description.

The Frohlich Hamiltonian of electron-phonon coupling system is\cite{fro}
\begin{eqnarray}
 & &H=\sum_{{\bf k},\sigma}(\epsilon_{\bf k}-\mu_0) d^{\dag}_{{\bf k},\sigma}
    d_{{\bf k},\sigma}
    +\sum_{\bf q}\omega_{\bf q} b^{\dag}_{\bf q}b_{\bf q} 
    +\frac{1}{\sqrt{N}}\sum_{\bf q}\sum_{{\bf k},\sigma}g_{\bf q}
    d^{\dag}_{{\bf k+q},\sigma}d_{{\bf k},\sigma}
    (b^{\dag}_{\bf -q}+b_{\bf q}) .
\end{eqnarray}
$N$ is the number of sites, $\epsilon_{\bf k}$ is
the bare band function. $b_{\bf q}$ and $d_{{\bf k},\sigma}$
are usual notations for phonon or electron operators, respectively.
$\mu_0$ is the chemical
potential, $\omega_{\bf q}$ the phonon frequency and $g_{\bf q}$ the
electron-phonon coupling. We set $\hbar=1$ and $k_B=1$.

Frohlich used a unitary transformation to treat $H$\cite{fro},
$H'=\exp (S)H \exp(-S)$.
The transformation can proceed order by order,
$$H'=H_0+H_1+[S,H_0]+[S,H_1]+{1\over 2}[S,[S,H_0]]+O(g^3_{\bf q}),$$
where $H=H_0+H_1$ and $H_0$ contains the first two terms and $H_1$
the last one. Frohlich let $H_1+[S,H_0]=0$ to get the generator $S$.
The transformation leads to a phonon-induced interaction with the
potential
\begin{eqnarray}
 & &V_F({\bf k+q,k})=\frac{g^2_{\bf q}\omega_{\bf q}}
    {|\epsilon_{\bf k+q}-\epsilon_{\bf k}|^2-\omega^2_{\bf q}},
\end{eqnarray}
which can be attarctive or repulsive with a singularity at
the energy shell: $|\epsilon_{\bf k+q}-\epsilon_{\bf k}|=\omega_{\bf q}$.
Frohlich noted that, although the transformation can eliminate the
first order terms completely, because of the singularity
one has to set a constraint $||\epsilon_{\bf k+q}-
\epsilon_{\bf k}|-\omega_{\bf q}|>\epsilon$ ($\epsilon>0$ is a constant)
in the transformation\cite{fro}. Because of the
constraint the elimination of the first-order terms is not conplete.

The BCS theory simplifies the Frohlich potential as\cite{bcs}
\begin{eqnarray}
 & &V_{BCS}({\bf k+q,k})=\left\{\begin{array}{ll}
    -g^2/\omega_p , & \mbox{for}
    |\epsilon_{\bf k+q}-\epsilon_{\bf k}|<\omega_p,\\
    0, & \mbox{otherwise}.
    \end{array}\right.
\end{eqnarray}

We propose to improve the Frohlich transformation by a new generator\cite{zh}
\begin{eqnarray}
 & &S={1\over \sqrt{N}}\sum_{\bf q}\sum_{{\bf k},\sigma}
    \frac{g_{\bf q}}{\omega_{\bf q}}
    d^{\dag}_{{\bf k+q},\sigma}d_{{\bf k},\sigma} 
    (b^{\dag}_{\bf -q}-b_{\bf q})\delta({\bf k+q,k}),
\end{eqnarray}
where $\delta({\bf k+q,k})$ is a function of the energies of incoming and
outgoing electrons in the electron-phonon scattering process,
\begin{eqnarray}
 & &\delta({\bf k+q,k})=\delta(\epsilon_{\bf k+q},\epsilon_{\bf k};
    \omega_{\bf q})=\left (1+\frac{|\epsilon_{\bf k+q}-\epsilon_{\bf k}|}
           {\omega_{\bf q}}\right )^{-1}.
\end{eqnarray}
The reason of choosing this generator will become clear lator.
After transformation the first order terms in $H'$ are
\begin{eqnarray}
 & &H_{I1}=H_1+[S,H_0]={1\over \sqrt{N}}\sum_{\bf q}\sum_{{\bf k},\sigma}
    \frac{g_{\bf q}}{\omega_{\bf q}+|\epsilon_{\bf k-q}-\epsilon_{\bf k}|}
    d^{\dag}_{{\bf k+q},\sigma} d_{{\bf k},\sigma} \nonumber\\
 & &\times\left\{[|\epsilon_{\bf k-q}-\epsilon_{\bf k}|
    -(\epsilon_{\bf k-q}-\epsilon_{\bf k})] b^{\dag}_{\bf -q}
    +[|\epsilon_{\bf k-q}-\epsilon_{\bf k}|
    +(\epsilon_{\bf k-q}-\epsilon_{\bf k})]b_{\bf q} \right\} .
\end{eqnarray}

The second order terms in $H'$ are:
\begin{eqnarray}
 & &H_{I2}=[S,H_1]+{1\over 2}[S,[S,H_0]] \nonumber\\
 & &={1\over 2N}\sum_{\bf k,q}\sum_{\sigma}
    \frac{g^2_{\bf q}}{\omega_{\bf q}}
    (b^{\dag}_{\bf -q}b_{\bf -q}+b_{\bf q}b^{\dag}_{\bf q})
    \delta^2({\bf k+q,k})\frac{\epsilon_{\bf k}-\epsilon_{\bf k+q}}
    {\omega_{\bf q}}\left ( d^{\dag}_{{\bf k+q},\sigma} d_{{\bf k+q},\sigma}
    -d^{\dag}_{{\bf k},\sigma} d_{{\bf k},\sigma} \right ) \nonumber\\
 & &-{1\over N}\sum_{\bf k,k',q}\sum_{\sigma,\sigma'}
    \frac{g^2_{\bf q}}{\omega_{\bf q}}
    \delta({\bf k+q,k})[2-\delta({\bf k'-q,k'})]
    d^{\dag}_{{\bf k+q},\sigma} d_{{\bf k},\sigma}
    d^{\dag}_{{\bf k'-q},\sigma'} d_{{\bf k'},\sigma'} ,
\end{eqnarray}
where the non-diagonal phonon terms are neglected. 
$H_{I2}$ contains a phonon-induced interaction with potential
\begin{eqnarray}
 & &V({\bf k+q,k})=-\frac{g^2_{\bf q}}{\omega_{\bf q}}
    \delta({\bf k+q,k})[2-\delta({\bf k,k+q})].
\end{eqnarray}
Fig.1 shows $V({\bf k',k})$ as functions of
$\epsilon_{\bf k'}-\epsilon_{\bf k}$ for different ratios $\omega_p/E_F$
where $E_F$ is the Fermi energy. For comparison we also show
$V_{BCS}({\bf k',k})$, which is a narrow square-well in the middle,
and the potential for small polarons which can be obtained via the
LF transformation: $V_{LF}=-g^2/\omega_p$ for all electrons.

The purpose of our transformation is 
to find a better way to divide the Hamiltonian into the unperturbed
part and the perturbation. After transformation
$H'=H'_0+H_{I1}+H'_{I2}+O(g^3_{\bf q})$.
The unperturbed part is
\begin{eqnarray}
 & &H'_0=\sum_{\bf q}\omega_{\bf q} b^{\dag}_{\bf q}b_{\bf q}
    +\sum_{{\bf k},\sigma}(E_{\bf k}-\mu_0)
    d^{\dag}_{{\bf k},\sigma}d_{{\bf k},\sigma} 
    -\sum_{{\bf k},\sigma}\Delta({\bf k})
    \left (d^{\dag}_{{\bf k},\uparrow}d^{\dag}_{{\bf -k},\downarrow}
    +d_{{\bf -k},\downarrow}d_{{\bf k},\uparrow} \right ),\\
 & &E_{\bf k}=\epsilon_{\bf k}+{1\over N}\sum_{\bf q}
    \frac{g^2_{\bf q}}{\omega_{\bf q}} \coth(\frac{\omega_{\bf q}}{2T})
    \frac{\epsilon_{\bf k+q}-\epsilon_{\bf k}}{\omega_{\bf q}}
    \delta^2({\bf k+q,k})\nonumber\\
 & &-{1\over N}\sum_{\bf q}\frac{g^2_{\bf q}}{\omega_{\bf q}}
    \delta({\bf k+q,k})[2-\delta({\bf k+q,k})]
    \frac{E_{\bf k+q}-\mu_0}{\xi_{\bf k+q}}
    \tanh\frac{\xi_{\bf k+q}}{2T} ,\\
 & &\Delta({\bf k})={1\over N}\sum_{\bf q}
    \frac{g^2_{\bf q}}{\omega_{\bf q}}
    \delta({\bf k+q,k})[2-\delta({\bf k+q,k})]
    \frac{\Delta({\bf k+q})}{\xi_{\bf k+q}}
    \tanh\frac{\xi_{\bf k+q}}{2T} ,\\
 & &\xi_{\bf k}=\sqrt{(E_{\bf k}-\mu_0)^2+\Delta^2({\bf k})},
\end{eqnarray}
which can be solved exactly. The perturbation is
$H_{I1}+H'_{I2}$ where $H'_{I2}=H_{I2}-(H'_0-H_0)$.
$E_{\bf k}$ and $\Delta({\bf k})$ have been determined by
the condition that the lowest order contribution of $H'_{I2}$ to the
self-energy is zero.

A renormalized chemical potential $\mu$ and a band renormalization
factor $\rho(\epsilon_{\bf k})$ can be introduced,
$E_{\bf k}-\mu_0=\rho(\epsilon_{\bf k})(\epsilon_{\bf k}-\mu)$.
The equation to determine $\mu$ is
\begin{eqnarray*}
 & &1-n=\frac{1}{N}\sum_{\bf k}\frac{\rho(\epsilon_{\bf k})
    (\epsilon_{\bf k}-\mu)} {\xi_{\bf k}}\tanh\frac{\xi_{\bf k}}{2T} ,
\end{eqnarray*}
where $n$ is the electron number density.
The equation to determine $\rho(\epsilon_{\bf k})$ will be given later.

The Green's function of $H'_0$ is\cite{mah}
\begin{eqnarray*}
 & &G_0({\bf k},\omega)=(\omega+E_{\bf k}-\mu_0)/(\omega^2-\xi^2({\bf k})),
    \mbox{~~}
    F^{\dag}_0({\bf k},\omega)=-\Delta({\bf k})/(\omega^2-\xi^2({\bf k})).
\end{eqnarray*}
The contribution of $H_{I1}$ to the self-energy (to the second
order of $g_{\bf q}$) iscite{mah}
\begin{eqnarray}
 & &\Sigma({\bf k},\omega)=-\frac{1}{N}\sum_{\bf q}\frac{g^2_{\bf q}}
    {(\omega_{\bf q}+|\epsilon_{\bf k-q}-\epsilon_{\bf k}|)^2}
    \frac{1}{\beta}\sum_{n}G_0({\bf k-q},\omega-i\omega_n)
    \nonumber\\
 & &\times \left\{
    \frac{[|\epsilon_{\bf k-q}-\epsilon_{\bf k}|
    -(\epsilon_{\bf k-q}-\epsilon_{\bf k})]^2}
    {i\omega_n-\omega_{\bf q}}
    -\frac{[|\epsilon_{\bf k-q}-\epsilon_{\bf k}|
    +(\epsilon_{\bf k-q}-\epsilon_{\bf k})]^2}
    {i\omega_n+\omega_{\bf q}}\right\},\\
 & &W({\bf k},\omega)=0.
\end{eqnarray}
The abnormal self-energy $W({\bf k},\omega)=0$ because it contains
a factor $|\epsilon_{\bf k-q}-\epsilon_{\bf k}|^2
-(\epsilon_{\bf k-q}-\epsilon_{\bf k})^2$ in the ${\bf q}$-summation.
Generally, the normal self-energy $\Sigma({\bf k},\omega)\neq 0$.
But for the normal state ($\Delta({\bf k})=0$) and when $T=0$,
we have
\begin{eqnarray}
 & &\Sigma_n(\epsilon_{\bf k}=\mu,\omega)=0,
\end{eqnarray}
that is, the normal self-energy is zero at the Fermi surface.
As the spectrum of elementary excitations in the normal state is
$\omega=\rho(\epsilon_{\bf k-q})(\epsilon_{\bf k-q}-\mu)
+\Sigma_n({\bf k},\omega)$,
the mass renormalization at Fermi surface is\cite{mah}
\begin{eqnarray}                                      
 & &\frac{m}{m^*}=\left [\rho(\epsilon_{\bf k})+
    \frac{\partial}{\partial \epsilon_{\bf k}}
    \Sigma_n({\bf k},\omega)\right ]\left /
    \left [1-\frac{\partial}{\partial \omega}
    \Sigma_n({\bf k},\omega)\right ]
    \right |_{\epsilon_{\bf k}=\mu}
    =\rho(\epsilon_{\bf k}=\mu) .
\end{eqnarray}
Eqs.(14), (15), and (16) are main reasons for
the choice of the functional form of $\delta({\bf k+q,k})$ in (5).
Now one can see clearly the purpose of our unitary transformation:
The transformed $H'$ is divided into the unperturbed part $H'_0$,
which contains the main physics of the problem, and the perturbation
$H_{I1}+H'_{I2}$, which is small as shown by Eqs.(14), (15), and (16).

Gap equation (11) can be rewritten as an integral equation by
introducing the Eliashberg function $\alpha^2 F$\cite{sca,grim}:
\begin{eqnarray}
 & &\Delta(\epsilon)=\int^D_{-D} d\epsilon'\int^{\infty}_{0}d\Omega
    \frac{N(\epsilon')}{N(\mu)}\alpha^2 F(\Omega)
    \frac{\Omega+2|\epsilon'-\epsilon|}{(\Omega+|\epsilon'-\epsilon|)^2}
    \frac{\Delta(\epsilon')}{\xi(\epsilon')}
    \tanh\frac{\xi(\epsilon')}{2T}.
\end{eqnarray}
$N(\epsilon)$ is the density of states (DOS).
The integration over $\epsilon'$ is from the bottom
($-D$) of the band to the top ($D$). The equation for
$\rho(\epsilon_{\bf k})$ can be rewritten in the same way,
\begin{eqnarray}
 & &\rho(\epsilon)=1+\int^D_{-D} d\epsilon'\int^{\infty}_{0}d\Omega
    \frac{N(\epsilon')}{N(\mu)}\alpha^2 F(\Omega)\coth(\frac{\Omega}{2T})
    \nonumber\\
 & &\times\left.\left\{\frac{\epsilon'-\epsilon}
    {(\Omega+|\epsilon'-\epsilon|)^2}
    -\frac{\epsilon'-\mu}{(\Omega+|\epsilon'-\mu|)^2}\right\}
    \right/ (\epsilon_{\bf k}-\mu) \nonumber\\
 & &-\int^D_{-D} d\epsilon'\int^{\infty}_{0}d\Omega
    \frac{N(\epsilon')}{N(\mu)}\alpha^2 F(\Omega)\left\{
    \frac{\Omega+2|\epsilon'-\epsilon|}{(\Omega+|\epsilon'-\epsilon|)^2}
    \right.\nonumber\\
 & &-\left.\frac{\Omega+2|\epsilon'-\mu|}{(\Omega+|\epsilon'-\mu|)^2}\right\}
    \frac{\rho(\epsilon')(\epsilon'-\mu)}{(\epsilon-\mu)\xi(\epsilon')}
    \tanh\frac{\xi(\epsilon')}{2T} .
\end{eqnarray}
These two equations are very similar to the Eliashberg equations\cite{elia,
sca,grim}.

For calculating the physical quantities we must calculate the thermodynamical
potential and the average of electron or phonon operators.
The thermodynamical potential is
\begin{eqnarray}
 & &\Omega=-{1\over \beta}\ln\mbox{Tr}\exp[-\beta H]  
    =-{1\over \beta}\ln\mbox{Tr}\exp[-\beta H'] 
    \approx -{1\over \beta}\ln\mbox{Tr}\exp[-\beta H'_0].
\end{eqnarray}
The last "$\approx$" is because of Eqs.(14) and (15), that is, 
to the order $O(g^2_{\bf q})$ the contribution of $H_{I1}+H'_{I2}$ is
very small in the lowest temperature region. Hence,
for the normal state 
\begin{eqnarray}
 & &\Omega\approx -{2\over \beta N}\sum_{\bf k}
    \ln\left\{1+\exp[-\beta \rho(\epsilon_{\bf k})
    (\epsilon_{\bf k}-\mu)]\right\}+\mbox{phonon part} .
\end{eqnarray}
The heat capacity can be calculated as
$C=-T\partial^2 \Omega/\partial T^2=C_0/\rho(\epsilon_{\bf k}=\mu)$,
where $C_0$ is the heat capacity for free electrons and
\begin{eqnarray}
 & &\rho(\epsilon_{\bf k}=\mu)=
    1-\lambda+\lambda\frac{\omega_p(\omega_p+D)}{(\omega_p+D)^2-\mu^2}.
\end{eqnarray}
In calculating we assume a constant DOS and $\lambda=2\int d\Omega
\alpha^2F(\Omega)/\Omega$. The enhancement factor at $\omega_p/D
\rightarrow 0$ is $1/(1-\lambda)$ which should be compared with
the same factor in ME theory $1+\lambda$\cite{grim}.

We have to take into account the effect of the unitary transformation
when calculating the thermodynamical average of electron or phonon
operators. The Green's function for the original Hamiltonian $H$
\begin{eqnarray}
 & &\tilde{G}({\bf k},\tau)=-\mbox{Tr}\left\{T_{\tau}\exp[-\beta (H-\Omega)]
    d_{{\bf k},\sigma}(\tau)d^{\dag}_{{\bf k},\sigma}\right\}
    \nonumber\\
 & &=-\mbox{Tr}\left\{T_{\tau}\exp[-\beta (H'-\Omega)]
    d'_{{\bf k},\sigma}(\tau)d^{\prime\dag}_{{\bf k},\sigma}\right\},
\end{eqnarray}
where $T_{\tau}$ means $\tau$ ordering\cite{mah}. 
The transformation of a single fermion operator can proceed as
$$d'_{{\bf k},\sigma}=e^{S} d_{{\bf k},\sigma} e^{-S}=d_{{\bf k},\sigma}+
    [S,d_{{\bf k},\sigma}]+{1\over 2}[S,[S,d_{{\bf k},\sigma}]]
    +O(g^3_{\bf q}).$$
\begin{eqnarray*}
 & &\tilde{G}({\bf k},\omega)=\int^{\beta}_0 d\tau
    \tilde{G}({\bf k},\tau)e^{\omega\tau} \nonumber\\
 & &=G({\bf k},\omega)-\frac{1}{N}\sum_{\bf q}\frac{g^2_{\bf q}}
    {\omega^2_{\bf q}}\delta^2({\bf k,k-q}) G_0({\bf k},\omega)
    \nonumber\\
 & &-\frac{1}{N}\sum_{\bf q}\frac{g^2_{\bf q}}
    {\omega^2_{\bf q}}\delta^2({\bf k,k-q}) 
    \frac{1}{\beta}\sum_{n}\frac{2\omega_{\bf q}}
    {(i\omega_n)^2-\omega^2_{\bf q}}G_0({\bf k-q},\omega-i\omega_n),
\end{eqnarray*}
where $G({\bf k},\omega)$ is the Green's function of $H'$.
Starting from the Green's function one can calculate various physical
quantities. As an example, we calculate the number of electrons
in a momentum state ${\bf k}$ for the normal state when $T=0$\cite{mah},
\begin{eqnarray*}
 & &n_{\bf k}=-{1\over \pi}\int^0_{-\infty} d\omega
    \mbox{Im}\tilde{G}({\bf k},\omega+i0^+)\nonumber\\
 & &=\left(1-\frac{1}{N}\sum_{\bf q}\frac{g^2_{\bf q}}
    {\omega^2_{\bf q}}\delta^2({\bf k,k-q})\right)
    \theta(\mu-\epsilon_{\bf k})
    +\frac{1}{N}\sum_{\bf q}\frac{g^2_{\bf q}}
    {\omega^2_{\bf q}}\delta^2({\bf k,k-q})
    \theta(\mu-\epsilon_{\bf k-q}).
\end{eqnarray*}
Fig.2 shows $n_{\bf k}$ as functions of $\epsilon_{\bf k}$ around the
Fermi surface. For smaller coupling (dashed line) or the larger frequency
(dash-dotted line), there is a finite jump of $n_{\bf k}$ at Fermi surface.
But for smaller frequency (solid line) and $lambda\sim 1$, the Fermi surface
is smeared by the electron-phonon coupling.
The jump of $n_{\bf k}$ at Fermi surface $\epsilon_{\bf k}=\mu$ is
$\rho(\epsilon_{\bf k}=\mu)$ (Eq.(25)) and 
it predicts an instability of the Fermi liquid state at
\begin{eqnarray}
 & &\lambda_c=\left(1-\frac{\omega_p(\omega_p+D)}{(\omega_p+D)^2-\mu^2}
    \right)^{-1}.
\end{eqnarray}
$\lambda_c\rightarrow 1$ when $\omega_p\rightarrow 0$.
For comparison, ME theory predicts a jump $1/(1+\lambda)$ and there is no
instability.

At the end, as the transformation is truncated after the second order of
$g_{\bf q}$, we justify the cutoff by showing the small expansion parameter.
Roughly speaking, in three-dimension it is $\lambda\omega_p/E_F$ when
$\omega_p/E_F< 1$ but $\lambda E_F/\omega_p$ when $\omega_p/E_F> 1$cite{zh}.
Note that they are the same as that of ME theory or LF transformation.

\newpage

\normalsize
\rm

\pagebreak

\begin{center}

\Large

{ \bf Figure Captions }

\end{center}

\vskip 0.5cm

\baselineskip 20pt

    {\bf Fig.1}~~~The phonon-induced interaction.

\vskip 0.5cm

    {\bf Fig.2}~~~The phonon-induced renormalization of the momentum
    distribution $n_{\bf k}$ on and near the Fermi surface.


\baselineskip 20pt

\begin{thebibliography}{99}

\bibitem{mig} A.B.Migdal, Zh. Eksp. Teor. Fiz. {\bf 34}, 1438 (1958) 
      [Sov. Phys. JETP {\bf 7}, 996 (1958)].

\bibitem{elia} G.M.Eliashberg, Zh. Eksp. Teor. Fiz. {\bf 38}, 966 (1960)
      [Sov. Phys. JETP {\bf 11}, 696 (1960)].

\bibitem{sca} D.J.Scalapino, The Electron-Phonon Interaction and
      Strong-Coupling Superconductors, in {\it Superconductivity}, edited
      by R.D.Parks (Makcel Dekker, New York, 1969).

\bibitem{grim} G.Grimvall, {\it The Electron-Phonon Interaction in
      Metals} (North-Holland, Amsterdam, 1981).

\bibitem{lf} I.G.Lang and Yu A.Firsov, Zh. Eksp. Teor. Fiz. {\bf 43},
      1843 (1962) [Sov. Phys. JETP {\bf 16}, 1301 (1963)].

\bibitem{hol} T.Holstein, Ann. Phys. (New York) {\bf 8}, 325 (1959).

\bibitem{am} A.S.Alexandrov and N.Mott, {\it Polarons and Bipolarons}
      (World Scientific, Singapore, 1995).

\bibitem{bcs} J.Bardeen, L.N.Cooper, and J.R.Schrieffer, Phys. Rev. {\bf 108},
      1175 (1957).

\bibitem{fro} H.Frohlich, Phys. Rev. {\bf 79}, 845 (1950); Proc. Roy. Soc.
      {\bf A215}, 291 (1952); Adv. Phys. {\bf 3}, 325 (1954).

\bibitem{zh} H.Zheng, M.Avignon, and K.H.Bennemann, Phys. Rev. {\bf B49},
      9763 (1994); H.Zheng, Phys. Rev. {\bf B50}, 6717 (1994);
      H.Zheng and S.Y.Zhu, Phys. Rev. {\bf B55}, 3803 (1997).

\bibitem{mah} G.D.Mahan, {\it Many-Particle Physics} (Plenum Press,
      New York, 1990).

\end{thebibliography}
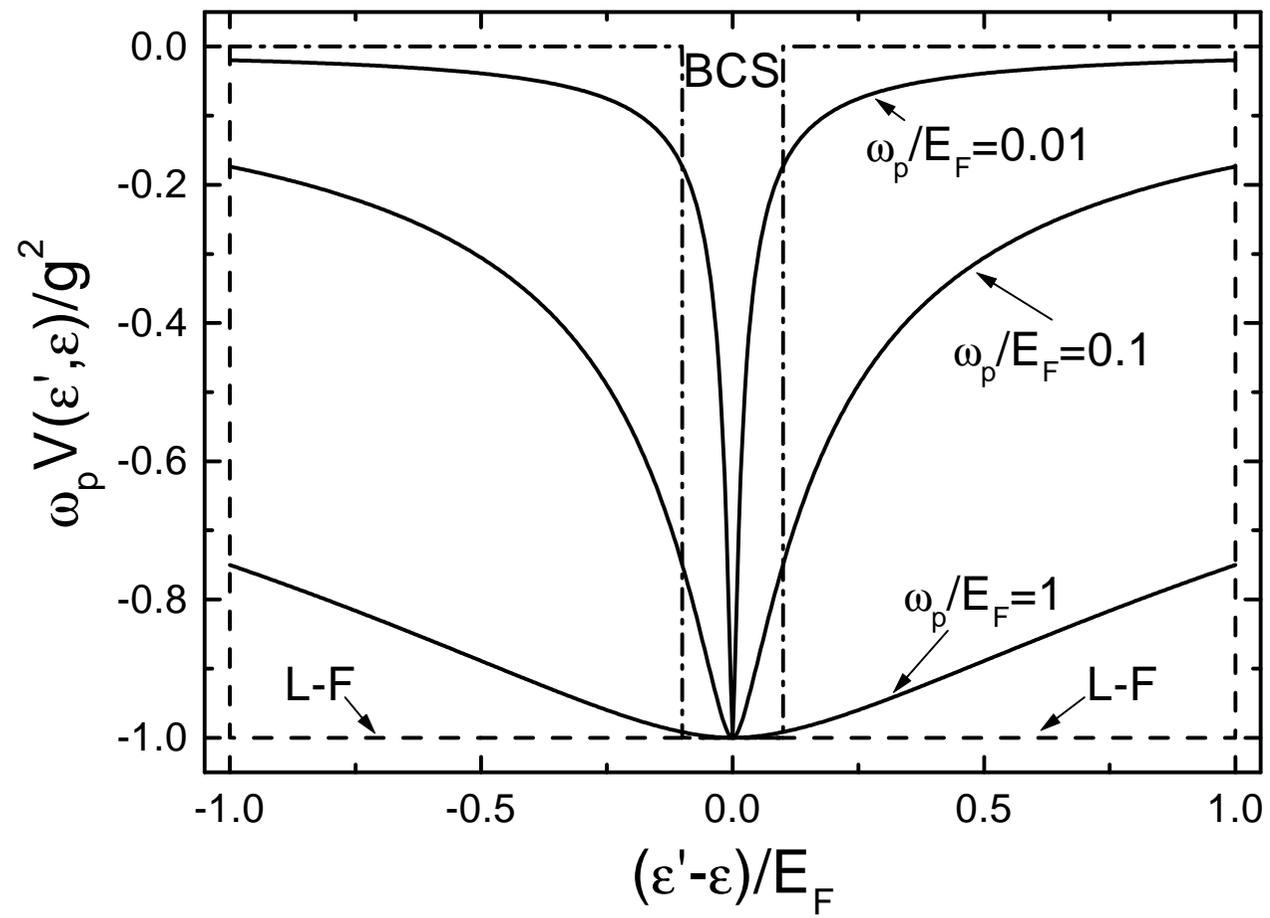
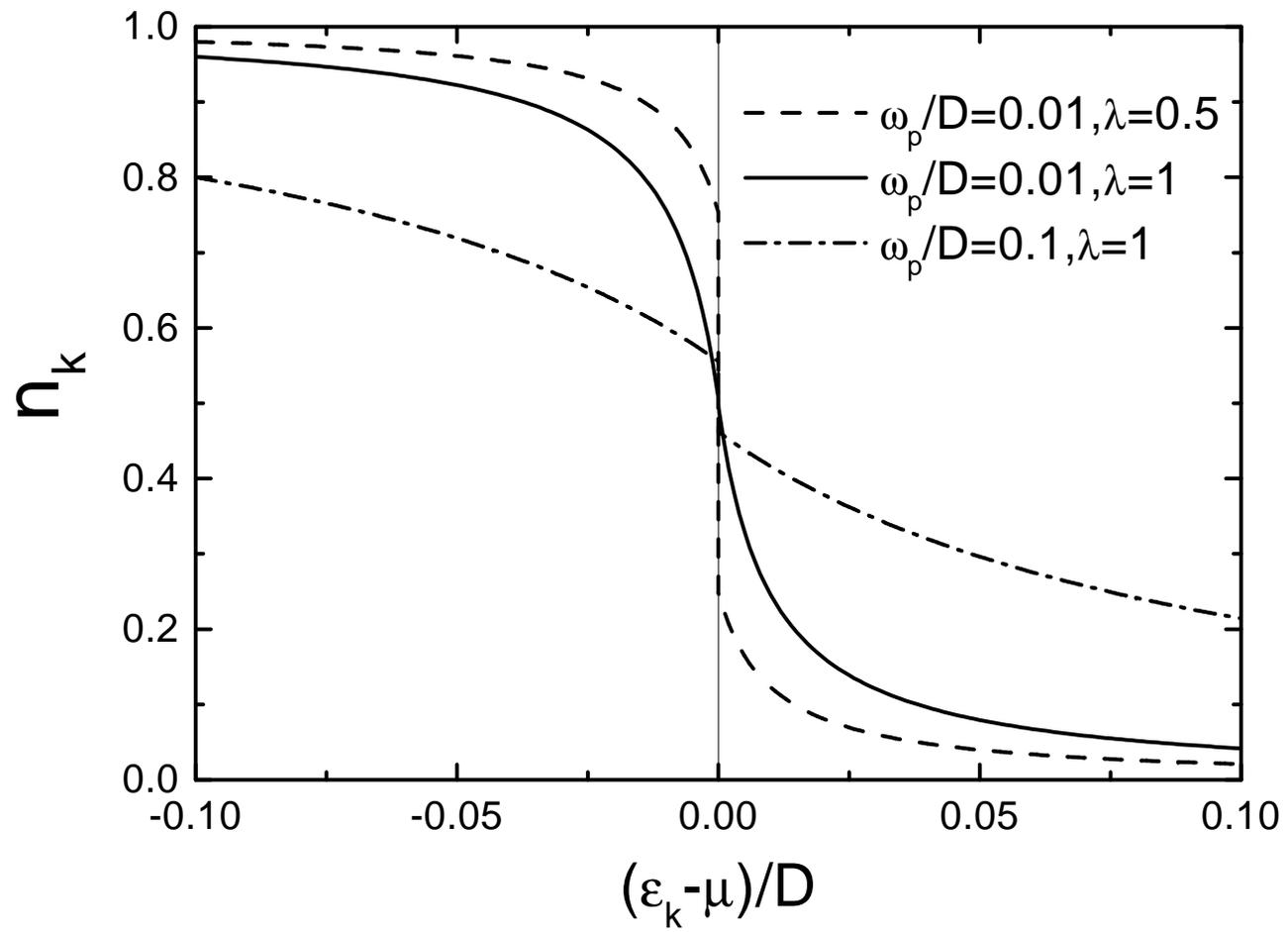
\end{document}